\documentclass[twocolumn]{aastex62}

\newcommand{\eg}{e.\,g.\ }

\newcommand{\ch}[1]{{\color{black} #1}}
\newcommand{\src}{NGC\,3115}
\newcommand{\msun}{M_\odot}

\shorttitle{}
\shortauthors{M. L. Jones et al.}

\begin{document}

\title{Investigating the Candidate Displaced Active Galactic Nucleus in \src}


\author{Megan L. Jones}
\affil{Department of Physics and Astronomy, West Virginia University, Morgantown, WV 26506, USA}
\affiliation{Center for Gravitational Waves and Cosmology, West Virginia University, Chestnut Ridge Research Building, Morgantown, WV 26505}

\author{Sarah Burke-Spolaor}
\affil{Department of Physics and Astronomy, West Virginia University, Morgantown, WV 26506, USA}
\affiliation{Center for Gravitational Waves and Cosmology, West Virginia University, Chestnut Ridge Research Building, Morgantown, WV 26505}

\author{Kristina Nyland}
\affiliation{National Radio Astronomy Observatory, 520 Edgemont Road, Charlottesville, VA 22903}

\author{Joan \ch{M.} Wrobel}
\affil{National Radio Astronomy Observatory, P.O. Box O, Socorro, NM 87801}

\begin{abstract}

The nearby galaxy \src\ contains a known radio-emitting, low-luminosity active galactic nucleus (AGN), and was recently claimed to host a candidate AGN displaced 14.3\,pc from the galaxy's optical photocenter. 
Our goal is to understand whether this represents a single offset AGN, an AGN in orbit around a central black hole, or something else.
We present a new, sensitive (RMS = {4.4}\,$\mu$Jy~beam$^{-1}$) 10\,GHz image, which finds evidence for only one AGN. We place a stringent limit on the radio luminosity of any secondary supermassive black hole of $L_{10\,GHz}<{5.8}\times10^{33}$ ergs/s. An analysis of the relative positioning of the radio core, X-ray nucleus, and stellar bulge in this galaxy indicate that the radio source is centrally located, and not offset from the galactic bulge. This provides an argument against a single offset AGN in \src, however does not provide conclusive evidence against the purported offset AGN as an in-spiralling secondary black hole.
\end{abstract}


\section{Introduction}

Supermassive black hole (SMBH) binaries should form during major galaxy mergers. 
Through the loss of orbital energy, the SMBH pair will eventually coalesce, releasing an enormous of amount of energy in the form of gravitational waves. Gravitational waves produced by SMBH mergers would be detectable by pulsar timing arrays \cite[e.g.][]{arz18}.
Asymmetric gravitational wave emission can produce a strong force---or ``kick"---to the final SMBH. 
If this kick is smaller than the host escape velocity, it can induce oscillation of the SMBH about the galaxy's core \citep[kick velocities can be up to several thousand km/s, causing the SMBH to be ejected from the galaxy;][]{ble16}. The SMBH should eventually settle into the host galaxy's center due to drag and other dynamical interactions with the stellar and gas environment \citep[\eg][]{begelman+80,campanelli+07}.
While recoiling SMBHs may return to the center after several Gyr, SMBH recoils in gas-poor galaxy mergers can remain non-centrally located for much longer periods of time \citep{ble16}. 

SMBH recoils induced by these kicks have astrophysical implications for the host galaxy, such as SMBH and galaxy evolution, galactic core structures, galaxy-SMBH
scaling relations, and the dependence of gravitational wave signals on redshift, among others \citep{kom12}.
The identification of potential recoiling SMBHs is therefore important in exploring past galaxy mergers, information on kick properties, as well as investigating predictions made via numerical relativity.

There are not many small-orbit binary SMBH systems known, with only a few examples below separations of 1\,kpc \citep[\eg][]{rod06}. There have likewise been scant discoveries of post-merger systems where the SMBH is seen in a state of offset/recoil, with only a few unconfirmed candidate systems \citep{kom08,pos12,ble13,len14,chi17}. To achieve sufficient AGN offsets such that the object is identifiable as an offset system, a large kick velocity is necessary. High kick velocities, while possible, are likely to cause stripping of much of the emissive material {(e.g. much of the narrow line region)} from the SMBH after its departure from the galactic center; therefore bright AGN with large offsets are likely to be rare. 

At a distance of 10.2 Mpc, \src\ is the nearest host of a billion-solar-mass black hole, 
and represents one of the first SMBHs with an accurate mass measurement based on stellar or gas dynamics (M$_{\rm{BH}} = 9.6 \times 10^{8}~M_{\odot}$)
\citep{korm92, gul09}. 
At the object's distance 1\arcsec = 49.5\,pc, making the task of spatially resolving any offset more feasible than for more distant sources.
Compact radio emission with a luminosity of $3.1 \times  10^{35}$ ergs/s that is coincident with the optical center in the nucleus of \src\ was first detected by \cite{wro12} by analyzing archival VLA data, with an Eddington luminosity of L$_{\rm{Edd}}$ = 1.2 $\times$ 10$^{47}$ ergs/s.
This detection is also coincident with an X-ray candidate nucleus 
identified by \cite{won11}, \ch{who conservatively estimate the luminosity as L$_X < 4.3 \times 10^{38}$ ergs/s}.
These data suggest the existence of a low-luminosity active galactic nucleus (AGN) residing in the center of this galaxy.

Using the 
Gemini-South telescope, \cite{men14} reported the detection of a broad-line H$\alpha$ emission \ch{with a luminosity of L$_{\rm{H\alpha}} = 4.2 \pm 0.4 \times 10^{37}$ ergs s$^{-1}$} that was
displaced from the photometric center of NGC~3115's stellar bulge by $290\pm50$ mas ($14.3\pm2.5$ pc). Upon inspecting several possibilities \ch{including a rotating relativistic disk around
the central black hole and imprecise starlight subtraction}, they concluded that the emission is most likely associated with an offset AGN. 
If this detection genuinely represents an offset AGN, there are two potential interpretations. First, it is possible the black hole fueling the offset AGN is actually in a pc-scale binary with a second black hole situated at the photocenter. Alternately, the AGN displacement could be the result of a black hole that has been kicked from the galaxy photocenter via recoil. 

In this paper we present a radio search for evidence of a \ch{binary} or offset AGN in \src. Section \ref{sec:data} reports new 10\,GHz data collected with National Science Foundation's Karl G. Jansky Very Large Array (VLA), Section~\ref{sec:radioresults} discusses the new radio measurements in the context of our detection of only one radio core, and reports a closer examination of the relative positions and astrometric errors of several different measurements of the AGN and galaxy center. We discuss the implication of these results in Section \ref{sec:discussion}.


\section{Very Large Array Data}\label{sec:data}

We observed \src\ with the VLA at X-band in the A-configuration on 12 June 2015, with 84 minutes on-target. The observational set-up had 64 frequency channels in \ch{each of} 32 unique spectral windows, across the range 7.976 GHz to 12.024 GHz with a center frequency of $\sim$10 GHz. The target pointing position was 10:05:13.927, --07:43:06.96.
We performed primary flux density and bandpass calibration using standard VLA calibrator 3C286, and used
J1007\ch{$-$}0207 as a phase calibrator. 

We calibrated the data using the VLA calibration pipeline, and interactively deconvolved the images using the {\sc clean} algorithm in the {\sc casa} software package {\citep{casa}}. 
The RMS of our final image is {4.4}~$\mu$Jy~beam$^{-1}$.
The synthesized beam had major and minor axes of 360 mas and 160 mas respectively, with a position angle of 40$^{\circ}$. 
{Multi-frequency synthesis was performed to account for the large fractional bandwidth with nterms=2 {due to the size of the imaged field}. We used briggs weighting with a robust value of 0.0 and a minpb of 0.2}. {The cellsize was set to 36 mas}.



\section{Analysis of Available Data}\label{sec:radioresults}

\begin{figure}
	\begin{center}			
		\includegraphics[trim=1cm 13cm 4cm 1cm, clip,width=0.99\columnwidth]{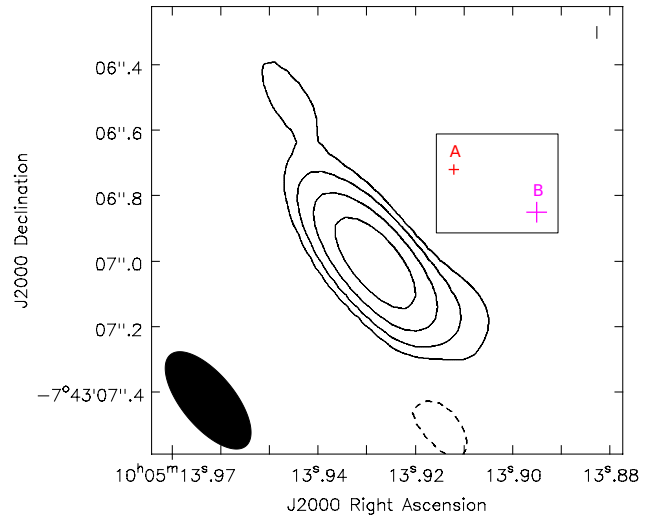}
	\caption{Contours of the 10-GHz emission from \src\ at -3, 3, 6, 12, and 24 times the RMS level in the image \ch{detected with the VLA}. The beam is displayed in the bottom left corner, \ch{with major and minor axes of 360 mas and 160 mas respectively and a position angle of 40$^{\circ}$. The noise level is 4.4~$\mu$Jy~beam$^{-1}$}. The relative positioning of the kinematic center (labelled ``B'') and the purported offset AGN (labelled ``A'') of \citet{men14} are represented by the purple and red crosses, respectively. Note that these are not absolute positions; they are displayed here to demonstrate that we should have been sensitive to two AGN if both were radio-emitting at the relative separation reported by \citet{men14}. The error bars on these two points represent the reported positional error for the bulge, and the AGN offset error for the AGN. {Negative contours are dashed and positive ones are solid.} 
	}
	\label{fig:image}		
	\end{center}
\end{figure}

\subsection{10\,GHz Measurement Results}

Our new radio image improved on the RMS sensitivity of our previous image by a factor of $\sim$5. Figure \ref{fig:image} shows 
a compact source located in the center of \src\ 
down to our detection limit of three times the RMS. The {\sc imfit} procedure in {\sc casa} was used to fit a two-dimensional elliptical-Gaussian to this sole source, yielding the integrated flux density, position, and one-dimensional position error appearing in Table 1. The tabulated position errors are reported as the radius of the error circle at the 95\% confidence level. The flux density error is the quadratic sum of the 3\% scale error \citep{per13} and the fit residual. The position error is the quadratic sum of terms due to the phase-calibrator position error (less than 2 mas), the phase-referencing strategies (estimated to be 100 mas), and the signal-to-noise ratio of the component (4 mas). The source was found to be point-like, with upper limits of {190} 
mas ({9.4} pc) on its major axis and {22} 
mas ({1.1} pc) on its minor axis, for a position angle of {$40.5\pm0.5$} degrees.

We detected two additional 10-GHz sources, each offset by more than
$1.5'$ from the nuclear 10-GHz source and thus unlikely to be associated
with it. As an independent cross-check of the VLA position error for the
nuclear 10-GHz source (Table 1), we searched the literature for
counterparts to the two offset sources.  Only one had any counterparts. 
Figure {2} shows that offset 10-GHz source plus the positions of its \textit{Chandra} X-ray
and $ugi$ counterparts
\citep{lin15,can15,can18}, and serves to validate the VLA astrometry.

\begin{figure}
    \centering
    \includegraphics[trim=10.5cm 4.4cm 10cm 3.8cm, clip, width=0.49\textwidth]{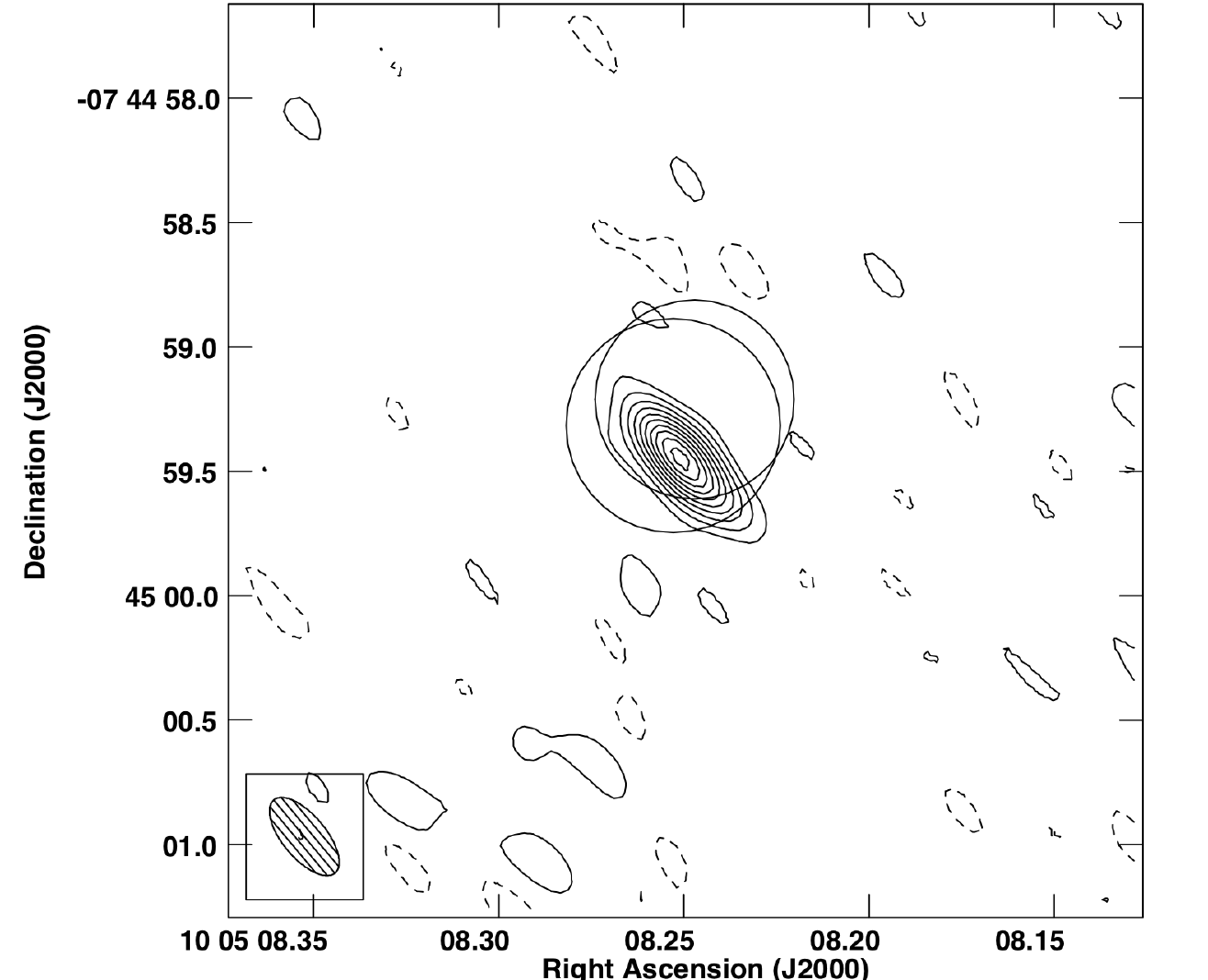}
    \caption{VLA image of Stokes $I$ emission near 10 GHz centered on a background source with a X-ray counterpart \citep{lin15} and an $ugi$ counterpart \citep{can15,can18}.  
    The symbols encode the counterpart positions and their errors at the 95\% confidence level. 
    {The slightly larger circle to the left represents the \textit{Chandra} X-ray counterpart, which is offset 147 mas from the 10-GHz background source. 
    The slightly smaller circle to the right represents the $ugi$ counterpart, with an offset of 249 mas from the background source.} For the 10-GHz image, the local rms noise is 11.4 $\mu$Jy~beam$^{-1}$ (1$\sigma$). The beam is displayed in the bottom left corner, with major and minor axes of 360 mas and 160 mas respectively and a position angle of 40$^{\circ}$. The allowed contours are at 1$\sigma$ times --6, --4, --2, 2, 4, 6, 8, 10, 12, 14, 16, 18, and 20.  Negative contours are dashed and positive ones are solid. 
    }
    \label{fig:stokes}
\end{figure}

\begin{table*}
\small
\begin{center}
\caption{Astrometric Position Comparison of Sources at Other Frequencies}
\label{tab:astrometry2}
\begin{tabular}{lllcccc}
\hline 
\hline
{\bf }  & \multicolumn{2}{c}{\bf Position, J2000} & {\bf \ch{Positional}} & {\bf Peak Flux Density} & {\bf Int. Flux Density}  &  {\bf Ref.}\\
{\bf Measurement}& {\bf RA} & {\bf Dec} & {\bf Uncertainty (mas)} & {\bf ($\mu$Jy beam$^{-1}$)} & {\bf ($\mu$Jy)} &  \\
	\hline	

2MASS & 10:05:13.93  & --07:43:07.1 & 120 &  --- & ---      & 1 \\

10 GHz   & 10:05:13.928 & --07:43:07.00 & 200 & 189 $\pm$ 5  & 207 $\pm$ 10 & 2 \\
8.5 GHz  & 10:05:13.927  & --07:43:06.96  & 200 &--- & 290 $\pm$ 30  & 3 \\
1.4 GHz  & 10:05:14.03   & --07:43:07.6 & 1000  & 400 $\pm$ 200 & --- &  4 \\
X-ray    & 10:05:13.93   & --07:43:07.0   &  460 & --- & ---     & 5 \\
\hline
M14 AGN      & 10:05:13.817   & --07:43:07.87  & --- & --- & --- & 6 \\
M14 Bulge    & 10:05:13.800   & --07:43:08.00  & --- & --- & --- & 6 \\
	\hline
	\hline
  \end{tabular}
  \end{center}
{\small {\bf Notes.} The 2MASS reference tie corresponds to the ICRS \citep{expsupp}.
We adopted the minor axis of the beam size for the error on our position measurement.
The \textit{Chandra} X-ray and \cite{men14} reference ties were obtained from comparing \textit{Chandra} and Gemini data to data from SDSS \citep{margutti+12}.
As the AGN found in \cite{men14} is measured to a position relative to the stellar kinematic center, we adopt the uncertainty on that measurement as the uncertainty in the position. 
\\
{\bf References.} (1) NED/2MASS; (2) this work; (3) \cite{wro12}; (4) \cite{whi97}; (5) \cite{eva10}; 
(6) \cite{men14}.\\
} 
\end{table*}

\begin{figure}
	\begin{center}			
		\includegraphics[trim=1.5cm 0.9cm 2.5cm 1.5cm, clip,width=0.99\columnwidth]{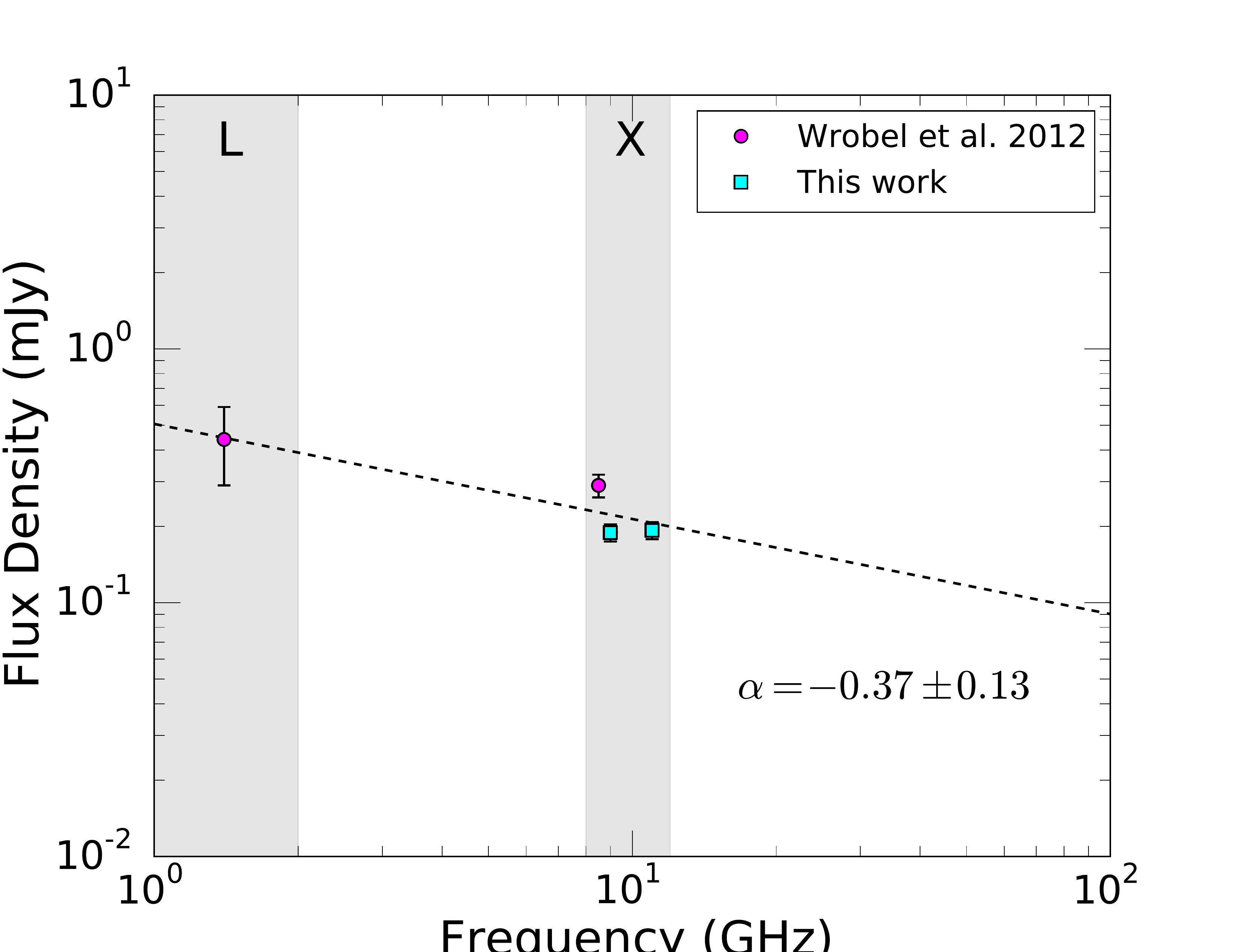}
  \leavevmode\smash{\makebox[0pt]{\hspace{-1em}
  \rotatebox[origin=l]{90}{\hspace{7.5em}
    Flux Density (mJy)}%
}}\hspace{0pt plus 1filll}\null

\vspace{-5mm}
Frequency (GHz)	
	\caption{Magenta circles represent data at 1.5 and 8.5 GHz from \ch{\cite{whi97} and} \cite{wro12}, \ch{respectively,} while the cyan squares signify data from this work, measuring the flux density after splitting the data into two sub-bands centered on 9 and 11 GHz. {The dashed line represents the best fit for all frequencies.}}
	\label{fig:spectrum}		
	\end{center}
\end{figure}

Using previous radio data as listed in Table \ref{tab:astrometry2} along with the data presented here, for the \src\ nuclear source we measure a spectral index of $\alpha={-0.37\pm0.13}$. This is consistent with the index measured by \cite{wro12} of $\alpha=-0.23\pm0.20$ (Fig.\,\ref{fig:spectrum}\ch{)}. 
The relatively flat spectral index of the emission indicates that this emission is likely related to a radio core component, i.e. marks the location of a SMBH rather than marking a distant jet outflow. 
\ch{The flat spectrum and persistence of the source show that it is likely in a
low-hard or quiescent state, consistent with the SMBH accreting slowly from the hot gas traced by the X-rays.}
\ch{The integrated flux measured in our new broadband VLA data from 8-12 GHz is consistent with that from archival VLA observations at 8.5 GHz (Table \ref{tab:astrometry2}).}
{The variability here cannot conclusively be determined}; thus, in the absence of \textit{strong} variability, there is no evidence for significant deviation from a single power-law in this radio component (i.e. we are not clearly observing a self-absorption turn-over, nor do we seem to be seeing two distinct regions with vastly different properties within our beam).

It is clear that we have not detected any radio source related to \src\ except for the one previously reported by \citet{wro12}. We initially set out to test the report of Menezes et al.\ that there was an active nucleus offset from the kinematic and photometric center of \src. As hypothesized in their paper, this could mean they either detected a single offset black hole, or an inspiralling black hole offset from another at the galaxy center, or that there is simply a single black hole at the galaxy center and the offset emission detected by Menezes et al.\ is caused by, for instance, an outflow.
However, several questions remain in the investigation of our initial hypotheses: First, \emph{should} we have detected a secondary black hole, given the detection of the first one? Second, can we determine whether this object corresponds to the galaxy photocenter or to the purported offset AGN of \citet{men14}? Third, might the central radio source actually encompass two SMBHs? The first question we address here, and the latter two require a discussion of the relative astrometry of measurements in other wavebands; this is discussed in Section \ref{sec:astrometry}.

\subsection{Would we have detected a distinct SMBH companion?}\label{sec:shouldwe}

The radio nucleus in this system has a low luminosity of 
$L_{\rm 10GHz}=8\times10^{34}$ ergs/s. 
In considering the hypothesis that this target could contain a \ch{binary} SMBH, we want to assess the probability that we should have detected a secondary SMBH in our observation given our limiting flux of three times the RMS of the off-source image, $S_{\rm lim}={13.8}~\mu$Jy beam$^{-1}$. At the distance of \src\ this corresponds to a limiting luminosity of 
$L_{\rm lim}={5.8}\times10^{33}$ ergs/s.
Due to the excellent VLA sensitivity and the small distance of this source, our limit on the luminosity of any secondary AGN is exceptionally low; in fact it lies several order of magnitudes below the span of published radio-quiet quasar distributions. If there is a companion, we would classify any secondary SMBH in this system as ``radio-silent'' \citep{padovani+15,padovani_agn_review}.

Past work has assessed the probability of finding a secondary AGN by integrating over radio luminosity functions down the limiting luminosity in an observation \citep[\eg][]{bur11}. If this target had a radio-loud or even radio-quiet secondary SMBH by the standard definitions, we should have detected it. It is possible that a secondary SMBH is not in an active state at all, in which case no waveband would have detected its emission. 
\ch{\cite{lut16} discuss gas patchiness as a result of stellar winds; the presence of a second SMBH accreting rapidly from a gas patch is also possible, which would show up as a broad-line AGN and only rarely exhibit radio emission.}

\begin{figure}
	\begin{center}			
		\includegraphics[width=1.0\columnwidth]{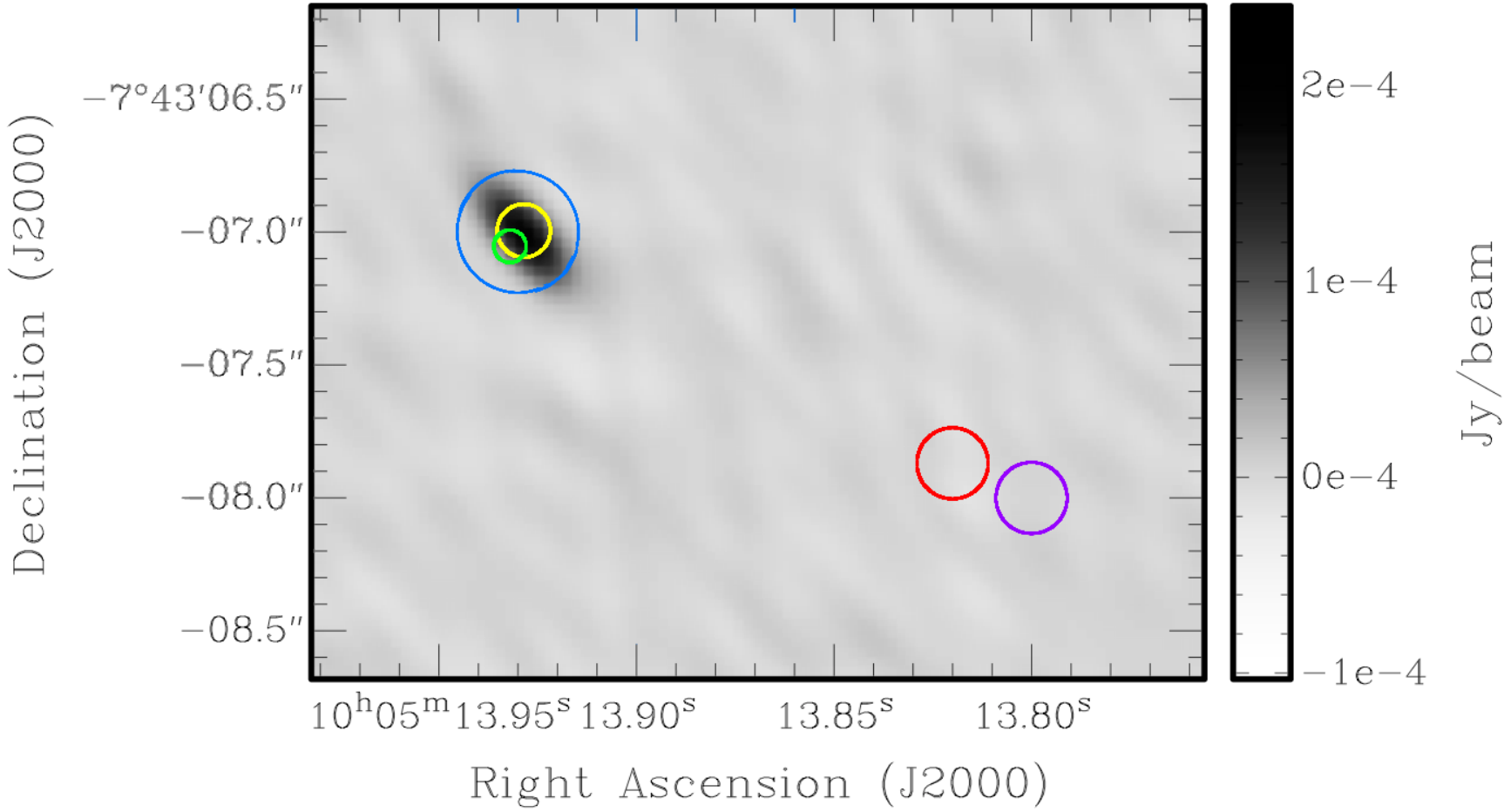}
	\caption{The relative positions of the data listed in Table \ref{tab:astrometry2}, with our 10\,GHz radio image shown in greyscale. The circle sizes represent the quadrature sum of the astrometric and measurement errors for each data point. The data are as follows: 2$\mu$m 2MASS centroid (green); X-ray core (blue); our radio position (yellow).
	\ch{The kinematic center measured by \citet{men14} is shown in purple (Menezes, private comm.).} Their reported offset AGN is shown in red, and is positioned in reference to the purple circle's position.
	}
	\label{fig:positions}		
	\end{center}
\end{figure}

\subsection{Multi-wavelength astrometry}\label{sec:astrometry}

Understanding the relative positions of the radio and X-ray emission, the purported offset AGN of \citet{men14}, and the galaxy kinematic center and/or photocenter is key to our interpretation of this object. The study of \citet{men14} benefited from precise position comparisons due to analysis of relative positioning within a single observation, while our study is by nature limited by astrometric and measurement errors. We will both examine astrometric errors, and re-examine the results of Menezes et al.

While \src\ is a well-studied object, there are relatively few works reporting on the actual position of the galaxy's kinematic center. 
Here we assess the optical photocenter, the X-ray component, and the radio component to understand what confidence we can have in their relative positioning. Unfortunately, there were insufficient numbers of objects detected both in our image and the X-ray/2MASS data to do a direct astrometric comparison. Instead, the reference tie and measurement errors reported by source publications for each observation are identified in Table~\ref{tab:astrometry2}.

The 8.5\,GHz observations were also performed with the VLA with the same frame tie accuracy \citep{wro12}, while the 1.4\,GHz measurement comes from the FIRST survey, which reports a mean astrometric precision of $50$\,mas \citep{whi97}. 
The 2$\mu$m photocenter position for \src\ was taken from 2MASS. The 2MASS survey is tied to the Tycho 2 catalog, which is accurate to $\sim$70\,mas \citep{expsupp}. The position and error listed is that for the \src\ photocenter from the 2MASS point source catalog \citep{2MASS}. 
The 0.5--7 keV \textit{Chandra} X-ray nuclear component was obtained from the {\it Chandra} Source Catalog \citep{eva10}. The localization of any potential AGN-related nucleus is limited by the fact that there is no nuclear point source, only a plateau of X-ray emission in the core of this galaxy \citep{won11}.
The X-ray reference tie is as quoted by \citep{eva10} for the 1$\sigma$ external astrometric error.
While the absolute astrometry of Gemini (used by \citealt{men14}) is about as accurate as that of {\it Chandra}, the astrometrically corrected position information was not reported by \citet{men14}. The AGN and bulge reported by \cite{men14} are relative positions in the frame of their observation, thus we do not show these errors in Fig.~\ref{fig:positions}.

\cite{men14} fit the 2D kinematic profiles of the data in order to obtain a position measurement for the offset broad H$\alpha$ emission line that indicates the presence of an AGN. They compare this position to the kinematic center of the galaxy (as determined by the velocity dispersions in that region) as well as the stellar bulge center (they equate the image of their collapsed data cube as representing the center of the stellar bulge) and determine that the AGN is coincident with neither the kinematic nor stellar bulge centers of \src.

In their analysis, Menezes et al.\ identified an offset of $\sim$290$\pm$50 mas ($\sim$14.3$\pm$2.5 pc), with the uncertainties determined using a Monte Carlo simulation. They determine that the kinematic center is coincident with the stellar bulge center, however do not report an absolute position of either the kinematic center, bulge center, or offset AGN. Based on observation headers that are not likely astrometrically corrected, the Gemini pointing center (directed at the stellar bulge center) was at a position of J2000 RA, Dec 10:05:13.80, --07:43:08.00 (R.~Menezes, private comm.).
The position of the off-centered AGN is reported as $\sim$260 mas east and 130 mas north from the stellar bulge center, giving an uncorrected AGN position of J2000 RA, Dec 10:05:13.82, --07:43:07.87.

The relative positions and net position errors are displayed atop 2MASS contours in Figure \ref{fig:positions}. The AGN detected by \citep{men14} is offset 1.84$''$ from our detected radio source, and their position measurement of the stellar kinematic center is 2.89$''$ from the 2MASS photocenter; thus there are clear residual absolute astrometry offsets in the observations of \cite{men14}.


\section{Discussion and Conclusions}\label{sec:discussion}

\subsection{Where is the radio AGN in \src?}

As shown in Figure \ref{fig:positions}, the position of our radio AGN agrees with the stellar bulge center as indicated by 2MASS, to significantly less than the separation between the offset between the bulge and AGN measured by Menezes et al. As such, it appears that the black hole related to this radio AGN is likely resident in, and not offset from, the galaxy center.
As seen in Figure \ref{fig:positions}, the position listed in the Menezes et al.\ {\sc fits} data file header do not appear to have any absolute astrometric correction applied. Regardless, if the bulge center were shifted to the 2MASS position, it is clear that the radio AGN is not colocated with the purported offset AGN.



\subsection{Does \src\ contain a binary, offset, or singular central AGN?}
Based on the relatively flat spectrum of the radio AGN and its colocation with the stellar bulge, it is clear that \src\ does not contain a singular offset (recoiling or wandering) AGN.

It is possible that the position-offset H$\alpha$ line does represent a separate SMBH, in which case the offset system may still be undergoing inspiral after a previous merger. Assuming this SMBH has a mass $\gtrsim10^8\,\msun$, the dynamical friction timescale of such an object at this separation is on the order of 100\,kyr, thus relatively short but not so infeasible to have detected it at this state of inspiral \citep{lac93}. The 2D kinematic measurements of Menezes et al.\ demonstrated some support for this argument, indicating that the isocontours of velocity dispersion demonstrated an elongated, elliptical shape, rather than showing a singular peak at the photometric center. However, the elongation was not along the axis of the AGN offset, suggesting some ongoing mobility of this secondary SMBH in the system under this hypothesis. There are no other large-scale indications that \src\ underwent a merger in the last few Gyr.

Finally, it is possible that \src\ simply contains a single, centrally located low-luminosity AGN that is giving rise to the radio source and some component of the central X-ray emission. If the offset source is indeed its own AGN, it appears to have no associated radio emission. 


\acknowledgments

MLJ and SBS 
are members of the NANOGrav Physics Frontiers Center which is supported by NSF award 1430284. SBS is supported by NSF EPSCoR award number 1458952. We thank R.~B. Menezes for providing helpful clarifications about the Gemini observations we discuss in this report. The National Radio Astronomy Observatory is a facility of the National Science Foundation operated under cooperative agreement by Associated Universities, Inc. 

\software{ CASA \citep{casa}}. %



\bibliography{ngc3115}{}



\end{document}